\documentclass[aps,prl,epsfigure,twocolumn,superscriptaddress]{revtex4-1}
\usepackage[colorlinks=true,linkcolor=blue,urlcolor=blue,citecolor=blue,pdfusetitle]{hyperref}
\usepackage[utf8]{inputenc}
\usepackage[english]{babel}
\usepackage{amsmath}
\usepackage[caption = false]{subfig}
\usepackage{graphicx,epstopdf}
\usepackage{blindtext}
\usepackage[table,xcdraw]{xcolor}
\usepackage{lipsum}
\usepackage{amsfonts}
\usepackage{bbm}
\usepackage{amssymb}
\usepackage{enumerate}
\usepackage{color}
\usepackage{latexsym}
\usepackage{physics}
\usepackage{times,txfonts}

\newcommand{\Acal}{\mathcal{A}}

\newcommand{\Ecal}{\mathcal{E}}
\newcommand{\Fcal}{\mathcal{F}}

\newcommand{\Lcal}{\mathcal{L}}

\newcommand{\Vcal}{\mathcal{V}}

\newcommand{\1}{\mathbbm{1}}

\newcommand{\Rmath}{\mathbbm{R}}

\newcommand{\interpro}[2]{\langle #1 | #2 \rangle}

\makeatletter
\newcommand\xleftrightarrow[2][]{%
	\ext@arrow 9999{\longleftrightarrowfill@}{#1}{#2}}
\newcommand\longleftrightarrowfill@{%
	\arrowfill@\leftarrow\relbar\rightarrow}
\makeatother

\begin{document}
	
	\title{Optimal charging of a superconducting quantum battery}
	
	\newcommand{\siqse}{Shenzhen Insititute for Quantum Science and Engineering, Southern University of Science and Technology, Shenzhen 518055, China}
	\newcommand{\physustech}{Department of Physics, Southern University of Science and Technology, Shenzhen 518055, China}
	\newcommand{\gdpkl}{Guangdong Provincial Key Laboratory of Quantum Science and Engineering, Southern University of Science and Technology, Shenzhen 518055, China}
	\newcommand{\szkl}{Shenzhen Key Laboratory of Quantum Science and Engineering, Southern University of Science and Technology, Shenzhen 518055, China}
	
	\author{Chang-Kang Hu}
	\affiliation{\siqse}\affiliation{\gdpkl}\affiliation{\szkl}
	
	\author{Jiawei Qiu}
	\affiliation{\siqse}\affiliation{\physustech}
	\author{Paulo J. P. Souza}
	\affiliation{Departamento de Física, Universidade Federal de São Carlos, Rodovia Washington Luís, km 235 - SP-310, 13565-905 São Carlos, SP, Brazil}
	
	\author{Jiahao Yuan}
	\affiliation{\siqse}\affiliation{\physustech}
	\author{Yuxuan Zhou}
	\affiliation{\siqse}\affiliation{\physustech}
	\author{Libo Zhang}
	\affiliation{\siqse}\affiliation{\gdpkl}\affiliation{\szkl}
	\author{Ji Chu}
	\affiliation{\siqse}
	\author{Xianchuang Pan}
	\affiliation{\siqse}
	
	\author{Ling Hu}
	\affiliation{\siqse}\affiliation{\gdpkl}\affiliation{\szkl}
	\author{Jian Li}
	\affiliation{\siqse}\affiliation{\gdpkl}\affiliation{\szkl}
	\author{Yuan Xu}
	\affiliation{\siqse}\affiliation{\gdpkl}\affiliation{\szkl}
	\author{Youpeng Zhong}
	\affiliation{\siqse}\affiliation{\gdpkl}\affiliation{\szkl}
	\author{Song Liu}
	\email{lius3@sustech.edu.cn}
	\affiliation{\siqse}\affiliation{\gdpkl}\affiliation{\szkl}
	\author{Fei Yan}
	\affiliation{\siqse}\affiliation{\gdpkl}\affiliation{\szkl}
	\author{Dian Tan}
	\email{tand@sustech.edu.cn}
	\affiliation{\siqse}\affiliation{\gdpkl}\affiliation{\szkl}

	\author{R. Bachelard} 
	\affiliation{Departamento de Física, Universidade Federal de São Carlos, Rodovia Washington Luís, km 235 - SP-310, 13565-905 São Carlos, SP, Brazil}
	\author{C. J. Villas-Boas} 
	\affiliation{Departamento de Física, Universidade Federal de São Carlos, Rodovia Washington Luís, km 235 - SP-310, 13565-905 São Carlos, SP, Brazil}
	
	\author{Alan C. Santos}
	\email{ac\_santos@df.ufscar.br}
	\affiliation{Departamento de Física, Universidade Federal de São Carlos, Rodovia Washington Luís, km 235 - SP-310, 13565-905 São Carlos, SP, Brazil}
	
	\author{Dapeng Yu}
	\affiliation{\siqse}\affiliation{\gdpkl}\affiliation{\szkl}\affiliation{\physustech}
	
	\begin{abstract}
		{Quantum batteries are miniature energy storage devices and play a very important role in quantum thermodynamics. In recent years, quantum batteries have been extensively studied, but limited in theoretical level.} Here we report the experimental realization of a quantum battery based on superconducting qubits. Our model explores dark and bright states to achieve stable and powerful charging processes, respectively. Our scheme makes use of the quantum adiabatic brachistochrone, which allows us to speed up the {battery ergotropy injection}.
		Due to the inherent interaction of the system with its surrounding, the battery exhibits a self-discharge, which is shown to be described by a supercapacitor-like self-discharging mechanism. Our results paves the way for proposals of new superconducting circuits able to store extractable work for further usage.
		
	\end{abstract}
	
	\maketitle
	
	\emph{Introduction --} In the past few decades, the miniaturization technology of integrated circuits has developed rapidly. In {their} 
	micro-structures, heat {exchanged with the environment} and {their own} quantum nature have begun to affect the function{ing} of devices, so one must start to consider the influence of quantum effect{s} on future machines~\cite{pekola2015towards, brandner2015thermodynamics, rossnagel2016single, halbertal2016nanoscale, partanen2016quantum, dutta2020single}. When we deal with these devices working in the quantum {regime}, 
	we need to carefully consider some thermodynamic concepts, such as work, heat, and entropy, which urge people to think about how to extend the laws of thermodynamics to quantum systems far away from the equilibrium state. On the other hand, the rapid development of quantum physics has made it possible to manufacture and precisely control large and complex quantum systems, such as trapped ions~\cite{leibfried2003quantum, wineland2013nobel, monroe2021programmable}, Bose-Einstein condensation~\cite{griffin1996bose, dalfovo1999theory}, superconducting circuits~\cite{you2011atomic, devoret2013superconducting}, and many others. Recent efforts {have been focused} on the study of a few particle {system} {and how some of its characteristics can be transformed into a statistical theory, thus }
	resulting in new macroscopic thermodynamic laws (such as Landauer's principle~\cite{landauer1961irreversibility}), giv{ing} rise to the new and exciting field of quantum thermodynamics~\cite{partovi1989quantum, vinjanampathy2016quantum, Alicki:18}.
	
	{Due} to the development of quantum thermodynamics, there is an {increasing} interest in developing new quantum devices {that are applied} to emerging quantum technologies, such as {those based on} quantum information processing, including components of quantum transistors and quantum diodes. In the context of quantum batteries (QBs), quantum phenomena, such as phase coherence~\cite{Alexia:20,Kamin:20-2} and entanglement~\cite{Ferraro:18,Crescente:20,Rossini:20,PRL_Andolina,Santos:20c}, constitute important resources, which can improve the performance of future scientific and technological equipment\textbf{s}. In this direction, {the development of storing quantum devices has been investigated in different systems~\cite{Andolina:19,Andolina:18,Andolina:19-2,Cruz:21}.}
	Yet the characterization of the charging process and energy storing performance due to the interactions with the environment is still largely unexplored~\cite{Gherardini:19,PRL2019Barra,Kamin:20-1}, even when no consumption center is connected to them~\cite{Santos:21b}. 
	
	\begin{figure}[t!]
		\centering
		\includegraphics[scale=0.45]{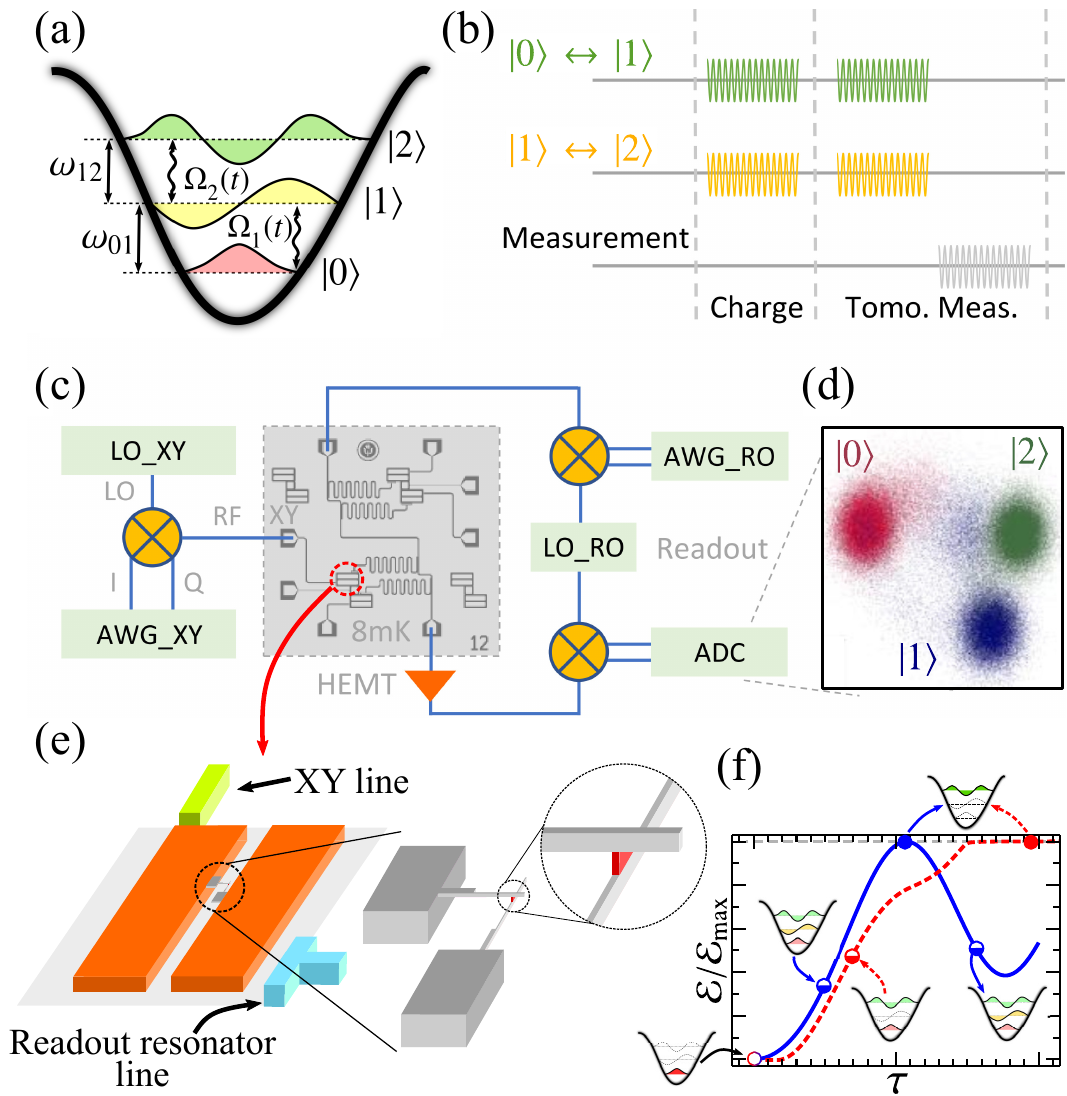}
		\caption{Schematic of the three-level quantum battery. (a) Three {lowest} levels of {a} transmon {define the} qutrit {system} used in our experiment. Two microwave fields are resonantly coupled to the $\ket{0}$ $\leftrightarrow$ $\ket{1}$ and $\ket{1}$ $\leftrightarrow$ $\ket{2}$ transitions to implement the charging processor. (b) Experimental {pulse} sequence. A tomography XY pulse and the resonant cavity readout pulse come after the charging pulse, to implement the qutrit's full state tomography. (c) Simplified diagram of the experimental setup. The transmon superconducting qubit sample is installed in the mixed chamber of a dilution cryostat to get the low temperature below 8 mK. The amplitude and phase controlled microwave pulses are generated with a quadrature IF (IQ) mixer, which is driven by a local oscillator (LO) and sideband modulated by a commercial arbitrary waveform generator (AWG). The readout pumping signal is generated in the same way and then is applied to the input port of the readout transmission line. {In the following,} the transmission signal is amplified by a high electron-mobility transistor (HEMT) amplifier, and is down-converted to the IQ signal, which will be digitized by an analog-to-digital converter (ADC) for further data analysis. (d) Single-shot readout performance measurements, three IQ clouds correspond to three energy level states. (e) The three dimension diagram of the transmon and the Josephson junction. (f) The schematic representation of a stable (dashed line) and unstable (continuum line) charging process. The choose of how we charge the battery is initially set in the external charging fields $\Omega_{1}(t)$ and $\Omega_{2}(t)$.
		}
		\label{Setup}
	\end{figure}

	In this paper we present the first experimental implementation of a QB, demonstrating {the} charging and self-discharging process of a {three-level} quantum cell {composed by a} superconducting device {with} {a superconducting transmon qutrit}. We {also} present a new approach to determine the optimal charging process based on quantum brachistocrone {for closed systems}~\cite{Rezakhani:09}, 
	{which allows us to determine the optimal interpolation functions for the time-dependent driving fields. Two distinct charging processes are implemented here:}
	Firstly, {a stable charge is achieved}, we make use of the stimulated Raman adiabatic passage (STIRAP), by controlling the time-dependent evolution of the {driving} field, {to} perform the adiabatic elimination {of an intermediate level} and bypass unwanted spontaneous discharge or attenuation~\cite{Santos:19-a,Santos:20c}. Secondly, by accurately controlling the evolution parameters and time of the driving field{s}, one enhances the charging power driving the QB through a fast and non-stable path with high fidelity. Finally, we investigate the loss of ergotropy of the battery due to its inevitable coupling to the environment (without any consumption center coupled to it), and show that, due to the particular decay rates of our system, it exhibits a super-capacitor behaviour.
	
	\emph{The superconducting device --} {As sketched in Fig.~\ref{Setup}{\color{blue}a}, we encoded our qutrit in the three lowest energy levels of the superconducting transmon {circuit}. The corresponding transition frequencies between the neighboring energy levels are $\omega_{01}\!=\!2\pi\times 6.266$~GHz and $\omega_{12}\!=\!2\pi\times 6.011$~GHz. The device energy level structure defines the QB energy levels from the bare Hamiltonian
		\begin{align}\label{H0}
			H_{0} = \sum\nolimits_{n=1}^{2} \hbar \omega_{(n-1)n} \ket{n}\bra{n} , 
		\end{align}
		$\ket{n}$ being the QB states with energy splittings $\epsilon_{n}\!=\!\hbar \omega_{(n-1)n}$, $\ket{0}$ the zeroth energy level. The relaxation and coherence times, extracted with standard state tomography measurements, are  $T_{01}^{relax}\!=\!19.4$~$\mu$s, $T_{12}^{relax}\!=\!12.5$~$\mu$s, $T_{01}^{coher}\!=\!26.7$~$\mu$s and $T_{12}^{coher}\!=\!9.9$~$\mu$s, respectively. We used two programmed microwave pulses, with time-dependent Rabi frequencies $\Omega_{1}(t)$ and $\Omega_{2}(t)$, to resonantly drive the qutrit and implement the expected charging time-dependent Hamiltonian. Then, to reconstruct the full density matrix of the qutrit, the standard quantum state tomography technology has been used, by performing a complete set of nine independent rotations between the charging and measurement pulses. Here, we rotate the quantum measurement bases {\color{blue}$\ket{\psi_{i}}$} to the ground state with the pulses {\color{blue}as shown} in Table.~\ref{tab_tomo}. The two microwave pulses are generated by analog IQ mixer down-conversion with a local oscillator with $\omega_{LO}\!=\!2\pi\times6.360$~GHz and AWG programmed microwave, and then they are applied to the qutrit through the XY control port, as shown in Fig.~\ref{Setup}{\color{blue}b} and ~\ref{Setup}{\color{blue}c}. The qutrit is capacitively coupled to a wandering readout resonator with a coupling strength of about $g_{r}\!=\! 2\pi\times23$~MHz. The readout frequency is $\omega_{r}\!=\!2\pi\times5.015$ GHz. The corresponding effective dispersive shift, $\chi\!=\!2\pi\times430$~KHz, is very close to the cavity linewidth, and meets the optimal dispersive readout conditions. {The IQ clouds corresponding to the three energy levels of our system are} shown in Fig.~\ref{Setup}{\color{blue}d}. More details about the superconducting circuit experimental setup can be found in the Supplementary Material~\cite{SupInf}.
	}
	
	\begin{table}[b]
		\centering  
		\scriptsize  
		\caption{{S}et of tomography measurement {bases} $|\psi_i\rangle$ and corresponding rotations $U_i$ sufficient to reconstruct {any arbitrary quantum density matrix of a qutrit}. }  
		\label{tab:notations} 
		\begin{ruledtabular}
			\begin{tabular}{rccp{0cm}}
				\hspace{0.5cm} {\bf \small i}  & {\bf\small $U_i$}  &  {\bf\small $|\psi_i\rangle$} &  \\  
				\hline 
				\hline\\[-2mm]  
				
				1 &       $I$                                               &  $\ket{0}$  & \\
				2 &       $(\pi)_x^{01}$                            &  $\ket{1}$  & \\
				3 &       $(\pi)_x^{01}(\pi)_x^{12}$        &  $\ket{2}$  & \\
				4 &       $(\pi/2)_y^{01}$                         &  $(\ket{0} -\ket{1})/\sqrt2$ & \\		
				5 &       $(\pi/2)_x^{01}$                         &  $(\ket{0} +j\ket{1})/\sqrt2$ & \\		
				6 &       $(\pi)_x^{01}(\pi/2)_y^{12}$     &  $(\ket{1} -\ket{2})/\sqrt2$ & \\
				7 &       $(\pi)_x^{01}(\pi/2)_x^{12}$     &  $(\ket{1} +j\ket{2})/\sqrt2$ & \\
				8 &       $(\pi/2)_x^{01}(\pi)_x^{12} $    &  $(\ket{0} -\ket{2})/\sqrt2$ & \\
				9 &       $(\pi/2)_y^{01}(\pi)_x^{12}$     &  $(\ket{0} +j\ket{2})/\sqrt2$ & \\
			\end{tabular}
		\end{ruledtabular}
		\label{tab_tomo}
	\end{table}
	
	\emph{Optimal charging process --} An optimal charging process takes into account both stability and charging speed. In our system, the energy is introduced in a stable way by employing {an adiabatic dynamics with} time-varying external fields to {inject} energy into the system{, in which the driving Hamiltonian reads
		\begin{align}
			H(t) =\hbar \Omega_{1}(t) (\ket{0}\bra{1}+\ket{1}\bra{0}) + \hbar\Omega_{2}(t) (\ket{1}\bra{2}+\ket{2}\bra{1}).
		\end{align}
		Due} to the adiabatic theorem validity conditions, the charging speed is negatively impacted leading to a loss of power (energy per time)~\cite{Santos:21c}. To bypass this issue, we explore the optimization process through adiabatic quantum brachistochrone (QAB)~\cite{Rezakhani:09} in order to speed up the QB ergotropy loading in context of adiabatic dynamics. The optimal trajectory is obtained through a variational formalism, where
	we find the set of differential equations for the {Rabi frequencies of the} driving fields $\Omega_{\ell}(t)$~(see~\cite{SupInf} for further details):
	\begin{subequations}
		\label{LagranTLS}
		\begin{align}
			( \Omega_{1}^2 + \Omega_{2}^2 ) \ddot{\Omega}_{1}^2 -2\left( 2{\Omega}_{2}\dot{\Omega}_{1}\dot{\Omega}_{2}+ {\Omega}_{1} (\dot{\Omega}_{1}^2 - \dot{\Omega}_{2}^2)\right) &= 0 , \\
			( \Omega_{1}^2 + \Omega_{2}^2 ) \ddot{\Omega}_{2}^2 -2\left( 2{\Omega}_{1}\dot{\Omega}_{1}\dot{\Omega}_{2} - {\Omega}_{2} (\dot{\Omega}_{1}^2 - \dot{\Omega}_{2}^2)\right) &= 0 ,
		\end{align}
	\end{subequations}
	where the boundary conditions depends on the kind of charging process. In order to achieve a stable drive, the system is driven through a dark state, which imposes $\Omega_{1}(0)\!=\!\Omega_{2}(\tau)\!=\!0$ and $(\Omega_{1}(\tau),\Omega_{2}(0))\!\neq\!(0,0)$, which corresponds to a \textit{stable} adiabatic charging~\cite{Santos:19-a}. Guaranteeing the stability of a QB is an important task to avoid backflow of charge from the QB to the charger, leading to a loss of efficiency when the external fields are not precisely controlled. On the other hand, as we shall see, when a high control of the charging fields is possible, the \textit{unstable} process can enhance the charging performance of the QB, which can be reached by setting $\Omega_{1}(\tau)\!=\!\Omega_{2}(0)\!=\!0$ and $(\Omega_{1}(0),\Omega_{2}(\tau))\!\neq\!(0,0)$.
	
	In scenarios in which the external driving fields present physical limitations (\textit{e.g.}, maximum available intensity or controllability), one needs to incorporate some constraints in the above equations. For example, our system has a single constraint associated with the maximum admissible amplitude for the independent fields $\Omega_{\ell}$ as given by $\Omega_{1}^2(t) + \Omega_{2}^2(t)\!\leq\!\Omega_{\text{max}}^2$, for all $t\in[0,\tau]$. Motivated by this limitation of a maximum field intensity, one can consider some specific classes of constraints, which lead to different solutions for Eqs.~\eqref{LagranTLS}. Since the strongest fields are desirable to achieve a fast adiabatic charging process, this leads to the following constraint: $\Omega_{1}^2(t) + \Omega_{2}^2(t)\!=\!\Omega_{\text{max}}^2$. From this assumption and considering the stable process, one obtains the solution
	\begin{align}
		\Omega_{1}^{\text{opt}}(t) &= \Omega_{\text{max}} \sin \left(\frac{\pi t}{2\tau}\right) , ~~ \Omega_{2}^{\text{opt}}(t)= \Omega_{\text{max}} \cos \left(\frac{\pi t}{2\tau}\right) . \label{ConstPower}
	\end{align}
	
	It is worth mentioning that since any adiabatic trajectory obtained here from Eqs.~\eqref{LagranTLS} comes from a first derivative of the adiabatic functional time, then there is no any information whether the solutions presented here are in fact optimal (minimum) curves. However, by using an approach via second derivative functional analysis (see~\cite{SupInf} for more details), we show that the above solution minimizes the functional time, so it is indeed the optimal adiabatic brachistochrone of the problem.
	
\begin{figure}[t!]
		\centering
		\includegraphics[scale=0.28]{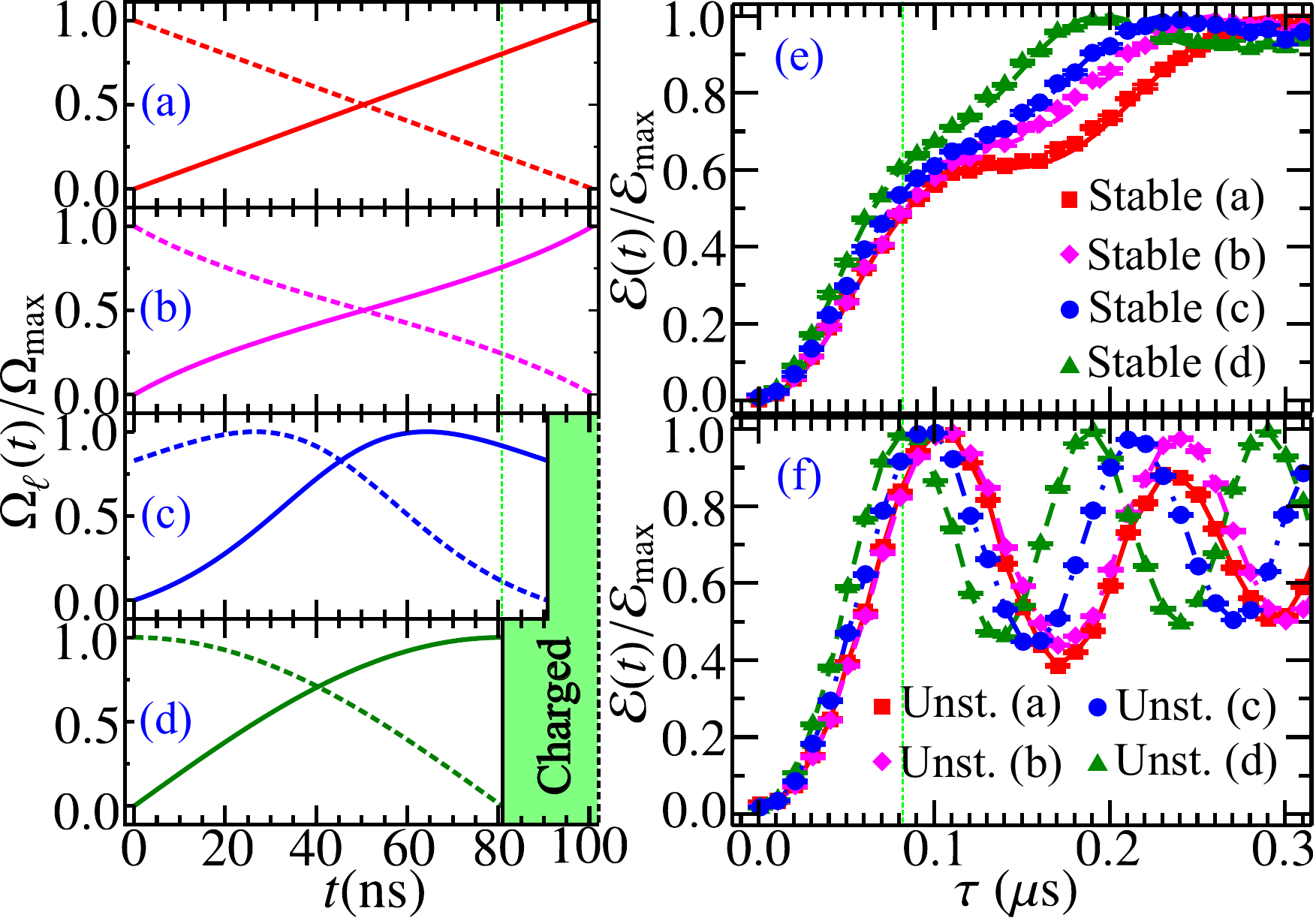}
		\caption{{\color{blue}(a--d)} Time dependence for the realized $\Omega_{1}$ (continuum line) and $\Omega_{2}$ (dashed line) for {\color{blue}(a)} a linear ramp, {\color{blue}(b)} linear and {\color{blue}(c)} quadratic constraints, and {\color{blue}(d)} no constraint. {\color{blue}(e--f)} Experimental result for the ergotropy as a function of the total evolution time $\tau$ (as a multiple of $\Ecal_{\text{max}}\!\approx\!51$~$\mu$eV) for {\color{blue}(e)} stable and {\color{blue}(f)} unstable charging process\textbf{es}. In the experiment we have set $\Omega_{\text{max}}\!=\!2\pi\times 10$~MHz.}
		\label{Fig-StablexUnstable}
	\end{figure}

	For the sake of comparison we consider the linear combination written as $\Omega_{1}(t) + \Omega_{2}(t)\!=\!\Omega_{\text{max}}$, whose brachistochrone solution is the arc of cycloid $\Omega_{1}^{\text{cyc}}(t) = \Omega_{\text{max}} \left[ 1-  \tan \left(  \pi\left(1-2t/\tau\right)/4\right)\right]/2$, and $\Omega_{2}^{\text{cyc}}(t)\!=\!\Omega_{\text{max}} - \Omega_{1}^{\text{cyc}}(t)$. {We also consider the case where the relation between the fields is set by the maximum admissible amplitude $\Omega_{1}^2(t) + \Omega_{2}^2(t)\!\leq\!\Omega_{\text{max}}^2$, {leading to a solution different from that given in Eq.~\eqref{ConstPower}.} In this situation,} Eq.~\eqref{LagranTLS} admits only numerical solutions. As a fourth case, we consider the linear ramp $\Omega_{1}^{\text{cyc}}(t)\!=\!\Omega_{\text{max}}(1-t/\tau)$ and $\Omega_{2}^{\text{cyc}}(t)\!=\!\Omega_{\text{max}}t/\tau$. The shape for each Rabi frequency implemented is shown in Figs.~\ref{Fig-StablexUnstable}{\color{blue}a}--\ref{Fig-StablexUnstable}{\color{blue}d} for the maximum power (which also corresponds to the minimum total evolution time for unstable dynamics).
	
	The charging process is realized by connecting the QB to the charger (time-dependent external fields). In a quantum battery, the stored charge is called ergotropy~\cite{Allahverdyan:04}, which quantifies the maximum amount of available work that can be extracted through unitary process\textbf{es}. It is given by $\Ecal(t)\!=\!\mbox{Tr}[\rho(t) H_{0}] - \min_{V\in \Vcal}\{\mbox{Tr}[V\rho(t) V^\dagger H_{0}]\}$, where the minimization is taken over the set $\Vcal$ of all unitary operators acting on the system. From the ordered set of eigenenergies $\epsilon_{1}\!\leq\!\epsilon_{2}\!\leq\!\cdots\!\leq\!\epsilon_{N}$, with eigenstates $\ket{n}$, of the internal battery Hamiltonian $H_0$ and the instantaneous spectral decomposition $\varrho_{1}(t)\!\geq\!\varrho_{2}(t)\!\geq\!\cdots\!\geq\!\varrho_{N}(t)$ of the instantaneous battery state $\rho(t)$, associated to eigenvectors $\ket{\varrho_{n}(t)}$~\cite{SupInf}, the ergotropy can be rewritten as
	\begin{align}
		\Ecal(t) = \sum\nolimits_{i,n}^{N,N} \varrho_{n}(t) \epsilon_{i} \left( |\interpro{\varrho_{n}(t)}{i}|^2 - \delta_{ni} \right) , \label{ErgotropyXstates}
	\end{align}	
	where we note that this definition is associated with a specific ordering of the eigenvalues of $\rho(t)$ and $H_0$, due to the $\delta_{ni}$ term. 

	Given the reference Hamiltonian of the system as defined in Eq.~\eqref{H0}, the instantaneous ergotropy for the stable process is shown in Fig.~\ref{Fig-StablexUnstable}{\color{blue}e}. The QB energy level spacing leads to a maximum storable ergotropy $\Ecal_{\text{max}}\!\approx\!51$~$\mu$eV. One can see that the QAB curve associated with the maximum intensity field provides the highest charging power for the QB. It means that, in a scenario in which physical constraints only impose a maximum value for the field intensity, the QAB allows us to get the optimal charging process by setting maximum power of the external fields at any instant of time. 
	
	One of the criteria for the high stability of a QB is its robustness against instantaneous discharging process~\cite{Santos:19-a}, which occur whenever systematic errors affect the external field control, leading then to an undesired energy leakage from battery. We now show how to enhance the QB performance, while keeping a high control to suppress energy backflow. To this end we change the field setup previously used, where now use a sequence in which $\Omega_{1}(\tau)\!=\!\Omega_{2}(0)\!=\!0$ and $(\Omega_{1}(0) , \Omega_{2}(\tau))\!\neq\!(0,0)$. The adiabatic stability is lost due to coherent superpositions of the eigenstates of the adiabatic Hamiltonian~(see~\cite{SupInf} for a detailed proof). Then, we compute the {adiabatic} brachistochrone for each case considered in Fig.~\ref{Fig-StablexUnstable}{\color{blue}a--d}, but now each $\Omega_{1}$ is represented by the dotted curve, while plain curves stand for $\Omega_{2}$. The instantaneous ergotropy is shown in Fig.~\ref{Fig-StablexUnstable}{\color{blue}f}. As a first result, one notes a power enhancement of more than $100\%$ of the unstable approach in comparison with the stable one for all cases considered here. In some cases, as in the ramp profile, this advantage is close to $300\%$. {In particular, it is worth highlighting the performance of the brachistochrone associated to the constant maximum power. Indeed, the performance of the QAB is the fastest in the 'stable' category ($\tau_{\text{c}}\!\approx\!190$~ns), approximately $100\%$ better than the unstable approach for the ramp case ($\tau_{\text{c}}\!\approx\!85$~ns).}
	In conclusion, the high control of the charging fields allows to suppress the instantaneous energy backflow: The adiabatic approach thus offers a greater stability, at the cost of speed.
	
\begin{figure}[t!]
		\centering
		\includegraphics[scale=0.35]{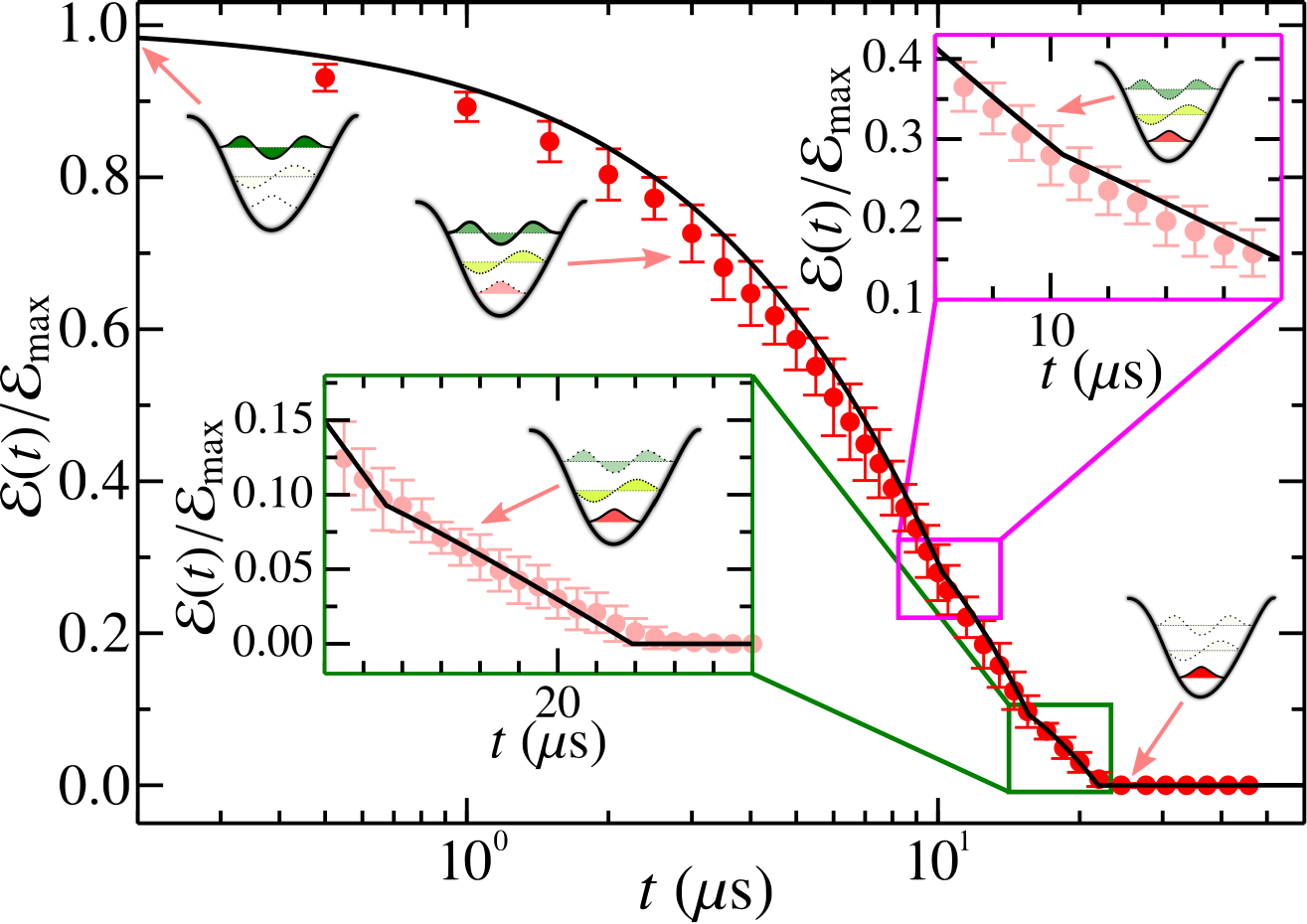}
		\caption{Instantaneous ergotropy (as a multiple of $\Ecal_{\text{max}}$) in the self-discharging process. The change of population is sketched in each step of the discharging. Inset we highlight the time interval in which the ergotropy changes due to the population inversion, as predicted from Eq.~~\eqref{Eq-Self-Discharging}. The decay rates are $\Gamma_{21}\!\approx\!51.4$~KHz and $\Gamma_{32}\!\approx\!79.7$~KHz.}
		\label{Fig-Self-Discharging}
	\end{figure}

	\emph{Self-discharging --} The characterization of the Transmon QB is completed by discussing its self-discharging behavior. The system is started in its fully charged state, without any external field or consumption center connected to it. The system is then governed by a cascade-like relaxation phenomenon which brings the system from its excited state to the ground state through a process in which state coherence is lost. The energy dissipation of the system is described by the master equation~\cite{Peterer:15}
	\begin{align}
		\dot{\rho}(t) =  \left[\frac{H_{0}}{i\hbar},\rho(t)\right] + \sum_{n,m}  \frac{\Gamma_{nm}}{2} \left[ 2 \sigma_{nm} \rho(t) \sigma_{mn} - \{\sigma_{nn},\rho(t)\} \right] ,
	\end{align}
	with $\sigma_{nm}\!=\!\ket{n}\bra{m}$, {the terms $\Gamma_{nn}$ and $\Gamma_{n(m\neq n)}$  being the dephasing rates and the crossed decay rates from state $\ket{n}$ to $\ket{m}$, respectively}. By solving the above equation we can analytically compute the instantaneous stored ergotropy decay's law as
	\begin{align}
		\Ecal(t) &= \left\{ \begin{matrix}
			\varepsilon_1 ( \varrho_1 - \varrho_0 ) + \varepsilon_2 ( \varrho_2 - \varrho_1 ) & \text{if}~\varrho_2\!>\!\varrho_1\!>\!\varrho_0 \\
			( \varrho_2 - \varrho_1 )(\varepsilon_2 - \varepsilon_1 ) & \text{if}~ \varrho_2\!>\!\varrho_0\!\geq\!\varrho_1 \\
			\varepsilon_2 ( \varrho_2 - \varrho_0 )& \text{if}~ \varrho_0\!>\!\varrho_2\!>\!\varrho_1 \\
			0 & \text{if}~ \varrho_0\!>\!\varrho_1\!>\!\varrho_2
		\end{matrix} \right. , \label{Eq-Self-Discharging}
	\end{align}
	with $\varepsilon_1\!=\!\epsilon_{1}$ and $\varepsilon_2\!=\!\epsilon_{1}+\epsilon_{2}$, and the quantities $\varrho_{n}\!=\!\varrho_{n}(t)$ {being} the instantaneous population in the battery state $\ket{n}$, given by
	\begin{align}
		\varrho_{1}(t) &= \frac{\Gamma_{21}}{\Gamma_{10}-\Gamma_{21}}\left[e^{-t\Gamma_{21}} - e^{-t\Gamma_{10}}\right] , ~~
		\varrho_{2}(t) = e^{-t\Gamma_{21}} \text{ , }
	\end{align}
	and $\varrho_{0}\!=\!1-\varrho_{1}(t)-\varrho_{2}(t)$. As an immediate consequence, there is a set of three crossing times $\tau^{\text{c}}_{n}$, which depend on the specific values of the relaxation rates $\Gamma_{21}$ and $\Gamma_{32}$. In fact, as highlighted in Eq.~\eqref{Eq-Self-Discharging} and sketched in Fig.~\ref{Fig-Self-Discharging}, the complete process is described by a sequence of incoherent population inversion, so that each time $\tau^{\text{c}}_{n}$ is obtained from conditions $\varrho_3\!=\!\varrho_2$, $\varrho_3\!=\!\varrho_1$ and $\varrho_2\!=\!\varrho_1$, respectively. Using a dynamics presented in Fig.~\eqref{Fig-StablexUnstable}, the system is initially prepared in a fully charged state with ergotropy $\Ecal_{\text{max}}\!\approx\!51$~$\mu$eV, then we let it decay and we compute its instantaneous ergotropy. The experimental decay curve is shown in Fig.~\ref{Fig-Self-Discharging}, where we highlight the time intervals in which the inversion population affects the ergotropy decay (see insets). 
	
	Differently from conventional classical batteries~\cite{Babu:20} and two-level QBs~\cite{Santos:21b}, which present an Ohmic decay behavior (i.e., characterized by a single time scale decay $\sim e^{-t/\tau_{\text{ohm}}}$~\cite{Babu:20,Santos:21b}), it is not possible to describe the self-discharging of transmon QBs using an Ohmic approach. In fact, the decay observed in the transmon QB needs to be explained through at least two mechanisms with two different time constants $\tau_{\text{1}}$ and $\tau_{\text{2}}$ as $\sim a_{1}e^{-t/\tau_{1}} + a_{2}e^{-t/\tau_{2}}$. Such a self-discharging behavior is the main {characteristic} of supercapacitors~\cite{Kowal:11,Lei-Zhang:18}. In the transmon QB, we understand this supercapacitor-like behavior as a consequence of both the multilevel nature of the system, and the asymmetry for the sequential decay rates $\Gamma_{21}$ and $\Gamma_{10}$.
	
	\emph{Conclusions --} In this paper we presented the first experimental realization of a quantum battery concerning the charging and self-discharging processes. By exploring the coherent coupling of a transmon three-level system with the external field, the battery can be coherently charged until its maximal charge. To enhance the battery charging speed while preserving its stability, the adiabatic process is optimized using the quantum brachistochrone theory, where the physical limitations of the external microwave {pulses} used to charge the battery are taken into account as constraints in the theory. Our results show that the optimal scheme to charge the battery comes from a joint adjustment of the kind of charging (stable or unstable) and the constraints on the fields used to inject energy in the system. The full characterization of a transmon three-level QB is studied by considering the self-discharging phenomena, which brings a fully charged battery into an empty charge final state. By computing the instantaneous ergotropy we verified that our transmon QB present a non-Ohmic discharging, therefore our single-cell QB is understood as a supercapacitor concerning such loss-energy phenomena. 
	
	{Our work constitutes a timely proposal of QB with potential applications in superconducting quantum computation, in which the stored work can be useful for quantum information processing, using, for example, the QB stored ergotropy to implement quantum gates. Our results are also applicable to any physical system, in which the dynamics of the three-level is driven by time-dependent external fields. The theory behind the optimal QAB studied here can also be adapted for other technologies that use three-level systems and STIRAP~\cite{Vitanov:17}, such as single-photon generation~\cite{Hijlkema:07}}.
	
	\emph{Acknowledgments --}
	This work was supported by the Key-Area Research and Development Program of Guang-Dong Province (Grant No. 2018B030326001), the National Natural Science Foundation of China (U1801661, 12004167, 11934010), the China Postdoctoral Science Foundation (Grant No. 2020M671861, {2021T140648}), the Guangdong Innovative and Entrepreneurial Research Team Program (2016ZT06D348), the Guangdong Provincial Key Laboratory (Grant No.2019B121203002), the Natural Science Foundation of Guangdong Province (2017B030308003), and the Science, Technology and Innovation Commission of Shenzhen Municipality (JCYJ20170412152620376, KYTDPT20181011104202253), and the NSF of Beijing (Grants No. Z190012). A.C.S., C.J.V.-B., and R.B. acknowledge the financial
	support of the São Paulo Research Foundation (FAPESP)
	(Grants No. 2018/15554-5, No. 2019/22685-1, No.
	2019/11999-5, and No. 2019/13143-0) and the Coordenação de Aperfeiçoamento
	de Pessoal de Nível Superior (CAPES/STINT), Grant No.
	88881.304807/2018-01. R.B. and C.J.V.-B. benefitted from
	the support of the National Council for Scientific and Technological Development (CNPq) Grants No. 302981/2017-9, No.
	409946/2018-4, and No. 307077/2018-7. C.J.V.-B. is also
	thankful for the support from the Brazilian National Institute
	of Science and Technology for Quantum Information (INCTIQ/CNPq) Grant No. 465469/2014-0.

	
%

\newpage

\onecolumngrid

\newpage

\begin{center}
	{\large{ {\bf Supplemental Material for: \\Optimal charging of a superconducting quantum battery}}}
	\vskip0.5\baselineskip{Chang-Kang Hu,$^{1,2,3}$ Jiawei Qiu,$^{1,4}$ Paulo J. P. Souza,$^{5}$ Jiahao Yuan,$^{1,4}$ Yuxuan Zhou,$^{1,4}$ Libo Zhang,$^{1,2,3}$ \\
	Ji Chu,$^{1}$ Xianchuang Pan,$^{1}$ Ling Hu,$^{1,2,3}$ Jian Li,$^{1,2,3}$ Yuan Xu,$^{1,2,3}$ Youpeng Zhong,$^{1,2,3}$ Song Liu,$^{1,2,3,{\color{blue}\ast}}$ \\
	Fei Yan,$^{1,2,3}$ Dian Tan,$^{1,2,3,{\color{blue}\dagger}}$ R. Bachelard,$^{5}$ C. J. Villas-Boas,$^{5}$ Alan C. Santos,$^{5,{\color{blue}\ddagger}}$ Dapeng Yu$^{1,2,3,4}$}
	
	\vskip0.5\baselineskip{{\em$^{1}$Shenzhen Insititute for Quantum Science and Engineering, \\Southern University of Science and Technology, Shenzhen 518055, China}\\
	{\em $^{2}$Guangdong Provincial Key Laboratory of Quantum Science and Engineering, \\Southern University of Science and Technology, Shenzhen 518055, China}
	\\
	{\em $^{3}$Shenzhen Key Laboratory of Quantum Science and Engineering, \\Southern University of Science and Technology, Shenzhen 518055, China}
	\\
	{\em $^{4}$Department of Physics, Southern University of Science and Technology, Shenzhen 518055, China}
	\\
	{\em $^{5}$Departamento de Física, Universidade Federal de São Carlos,\\ Rodovia Washington Luís, km 235 - SP-310, 13565-905 São Carlos, SP, Brazil}
	 }
	\vskip0.5\baselineskip{$^{\color{blue}\ast}$lius3@sustech.edu.cn, ~~~ $^{\color{blue}\dagger}$tand@sustech.edu.cn , ~~~ $^{\color{blue}\ddagger}$ac\_santos@df.ufscar.br}
\end{center}




\twocolumngrid

\appendix

\setcounter{equation}{0}
\setcounter{figure}{0}
\setcounter{table}{0}

\renewcommand{\theequation}{S\arabic{equation}}
\renewcommand{\thefigure}{S\arabic{figure}}

\section{Experimental setup}

\begin{figure*}[t!]
	\centering
	\includegraphics[scale=0.8]{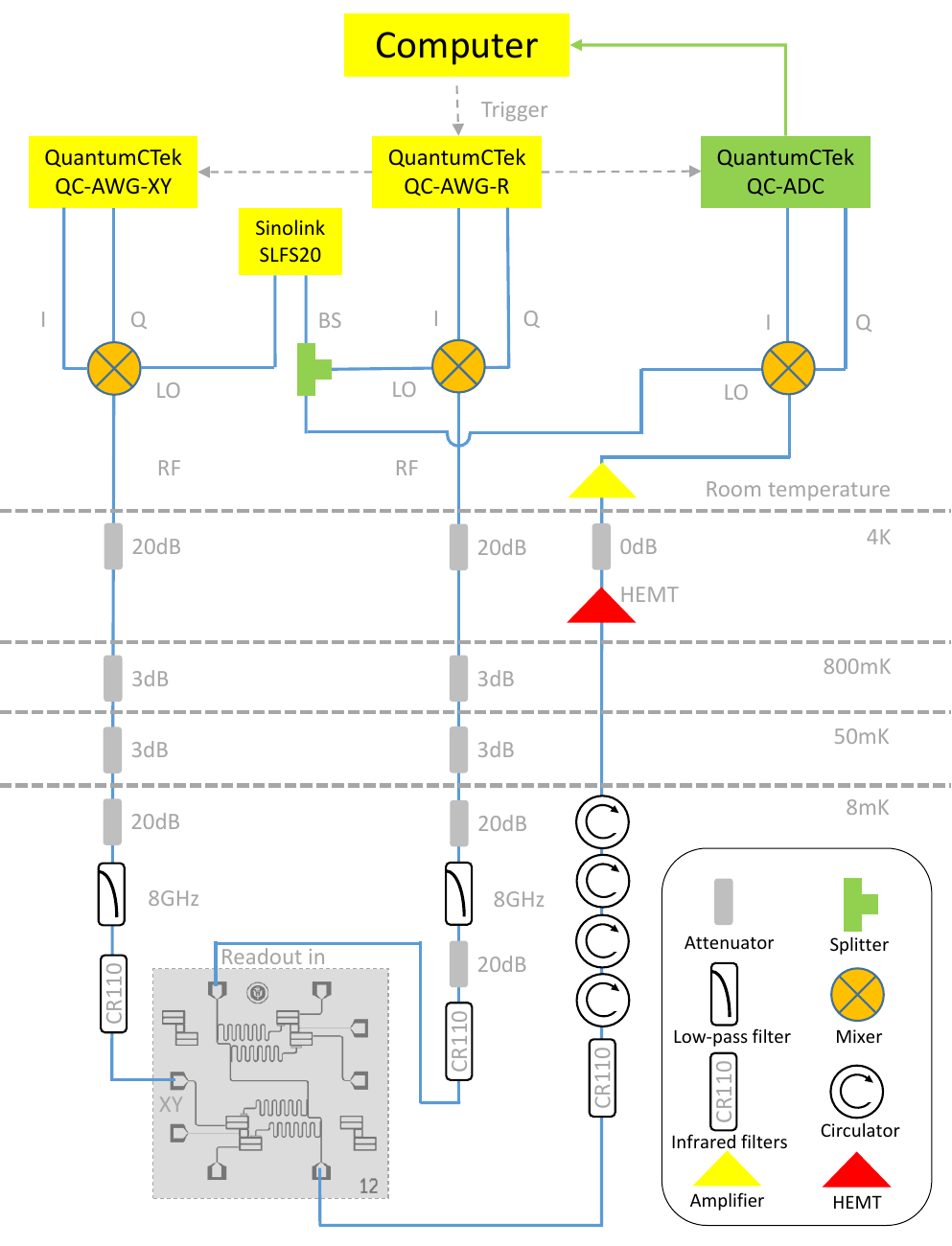}
	\caption{Electronics, cryogenics, and sample schematic of our experimental setup}
	\label{Ap-Figure_Setup-SM}
\end{figure*}

The transmon superconducting qutrit device is installed inside a dilution refrigerator system and cooled down to under 8 mK. Electronics, cryogenics, and sample diagram {{are}} shown in Fig.~\ref{Ap-Figure_Setup-SM}{\color{blue}}. The qubit drive pulses are generated with a quadrature IF (IQ) mixer. The local oscillator (LO) is supported by a commercial Mulitichannel coherence microwave generator Sinolink SLFS20. The IQ signals are generated by a QuantumCTek arbitrary waveform generator (AWG), with a 2 GHz sampling rate. The IQ signal amplitude is directly proportional to Rabi frequency, while the control signal phase determines the rotation axis. So, we could program the AWG to implement our targeted Hamiltonian. The readout pumping signal is generated in the same way and then is applied to the input port of the readout transmission line. After being amplified by a high electron-mobility transistor (HEMT) amplifier at the 4K stage and a low noise room temperature amplifier, {{the}} readout signal {{is}} downconverted to the IQ signals with the same LO microwave source. Finally, the demodulated IQ signals will be digitized by analog-to-digital converters (ADC).

\section{The Brachistochorne}

By using the definition of the Lagrangian in Ref.~\cite{Rezakhani:09} we {{identify}} the Lagrangian for our quantum battery charging process {{as}} 
\begin{equation}\label{Ap-eq:qutritLagrangian}
\mathcal{L}(\Omega, \dot{\Omega})= \left(\frac{|| \dot{H}(t)||}{\Delta^2(t) }\right)^2 = \frac{\dot{\Omega}_{1}^2(t) + \dot{\Omega}_{2}^2(t)}{\left( \Omega_{1}^2(t) + \Omega_{2}^2(t)\right)^2}  ,
\end{equation}
where $\Omega = (\Omega_1, \Omega_2)$, dot denotes $d/dt$, $||A||=\sqrt{\mathrm{Tr}(A^\dag A)}$ is the Frobenius norm, $\Delta(t)$ is the (minimum) instantaneous energy gap among all the eigenstates of $H(t)$.
Further, after applying the Euler-Lagrange equations
\begin{align}
\frac{\partial \Lcal (\Omega, \dot{\Omega})}{\partial \Omega_{n}} - \frac{d}{dt} \left(\frac{\partial \Lcal (\Omega, \dot{\Omega})}{\partial \dot{\Omega}_{n}} \right) = 0, ~~ n\in\{1,2\} ,
\end{align}
we find the system of coupled second order differential equations,
\begin{equation}\label{Ap-eq:EL_unconstrained}
\begin{split}
(\Omega_1^2 + \Omega_2^2)\ddot{\Omega}_1 - 2\left(2\Omega_2\dot{\Omega_1}\dot{\Omega_2} + \Omega_1(\dot{\Omega}_1^2- \dot{\Omega}_2^2)\right) &= 0\text{, }\\
(\Omega_1^2 + \Omega_2^2)\ddot{\Omega}_2 - 2\left(2\Omega_2\dot{\Omega_1}\dot{\Omega_2} - \Omega_2(\dot{\Omega}_1^2- \dot{\Omega}_2^2)\right) &= 0\text{, }
\end{split}
\end{equation}
where $\Omega_{1}$ and $\Omega_{2}$ are expressed in function of $t$. 
In the case where the above equation does not admit a analytical solution, we numerically solve this system of differential equations using the algorithms in the SciPy library~\cite{2020SciPy-NMeth}.

\subsection{Adiabatic interpolations}

In general, it is not possible to find the analytical solution of Eq.~(\ref{Ap-eq:EL_unconstrained}).
Nonetheless, we can impose constraints to the canonical variables, $\Omega_1(t)$ and $\Omega_2(t)$, {{which allows for an}} analytical treatment. Firstly, we can impose the linear constraint, $\Omega_{1}(t) + \Omega_{2}(t)= \Omega_{\text{max}}$. So, from Eq.~\eqref{Ap-eq:qutritLagrangian} and imposing the boundary conditions $\Omega_1(0)\!=\!\Omega_{\text{max}}$ and $\Omega_1(\tau)\!=\!0$ of the stable process, we get the solution for the Euler-Lagrange equation as
\begin{equation}
\Omega_{1}^{\text{sta}}(t) = \Omega_{\text{max}}\left\{\frac{1}{2} - \frac{1}{2}\tan\left[\frac{\pi \left(1-2t/\tau\right)}{4}\right]\right\}  .
\end{equation}

Conversely, in the case where the charging process is unstable, we have the conditions $\Omega_1(0)\!=\!0$ and $\Omega_1(\tau)\!=\!\Omega_{\text{max}}$ and the solution is then
\begin{equation}
\Omega_{1}^{\text{uns}}(t) = \Omega_{\text{max}}\left\{\frac{1}{2} + \frac{1}{2}\tan\left[\frac{\pi \left(1-2t/\tau\right)}{4}\right]\right\}  .
\end{equation}

Now, we consider the class of constraints in which $\Omega_{1}^2(t) + \Omega_{2}^2(t) = \Omega_{\text{max}}^2$, so that the Euler-Lagrange equation provides
\begin{equation}\label{Ap-eq:quadraticELE}
\ddot{\Omega}_{1} \left(\Omega_{1}^2 - \Omega_{\text{max}}^2 \right) - \Omega_{1} \dot{\Omega}^{2}_{1} = 0 ,
\end{equation}
whose the solution for the stable process reads
\begin{equation}\label{Ap-eq:quadratic_sol}
\Omega_{1}(t)=\Omega_{\text{max}}\sin\left(\frac{\pi t}{2\tau}\right) .
\end{equation}

It is possible to show that for the unstable charging we have $\Omega_{1}(t)\!=\!\Omega_{\text{max}}\cos(\pi t/(2\tau))$.

\section{The second Derivative test}

Let us {{start}} by considering the functional $A[q]$ given by the equation
\begin{align}
A[q] = \int_{t_0}^{\tau} \Lcal(q,\dot{q}) dt  .
\end{align}
We are interested in obtaining the critical ``points" for $A[q]$. Since the generalized coordinates and velocities, $q$ and $\dot{q}$, define a vector in phase space of the system, it is reasonable to assume that the elements $q$ of the vector space $\Fcal$ satisfy
\begin{align}
A = \{ q : [t_{0},\tau] \rightarrow \Rmath^n~|~q_{n} \in C^2, q(t_{0})=q_{0}, q(\tau)=q_{\tau} \},
\end{align}
{{where $C^2$ is the class of twice-differentiable functions.}}
The physical meaning of the above equation is that the position and momentum of a particle are functions with well-defined mathematical behavior. As we shall see, the such condition for the QAB means the external fields that act on the system are experimentally feasible.
Because $A[q]$ is defined from a vector space, the derivative of $A[q]$ is defined from the Gateaux's derivative
\begin{align}
D_{\eta} A[q] = \left. \frac{d A[q+\epsilon\eta]}{d\epsilon} \right\vert_{\epsilon=0} = \lim_{\epsilon\rightarrow 0} \frac{A[q+\epsilon\eta] - A[q]}{\epsilon} , \label{Ap-EqDeta}
\end{align}
where $\epsilon\in \Rmath$ and the parameter $\eta\in\Fcal$, satisfying the boundary conditions $\eta(t_0)\!=\!\eta(\tau)\!=\!0$, are the admissible $\eta$'s. Because $\eta$ {{is a vector}}, the derivative in Eq.~\eqref{Ap-EqDeta} {{is}} also known as directional derivative along direction $\eta$. The condition $\eta(t_0)\!=\!\eta(\tau)\!=\!0$ needs to be satisfied in order to guarantee that the solutions will keep the initial and final conditions of the physical system. Again, in the QAB case such conditions are associated with initial and final values of the fields used to drive the system. Then, after some calculations we find
\begin{align}
D_{\eta} A[q] = \sum_{j=1}^{n}\int_{t_0}^{\tau} \left[ \frac{\partial \Lcal}{\partial q_{j}} - \frac{d}{dt}\left(\frac{\partial \Lcal}{\partial \dot{q}_{j}}\right) \right]\eta_{j} dt ,
\end{align}
where $q_{j}$ (resp. $\dot{q}_{j}$ and $\eta_{j}$) denotes the $j$-th component of the generalized coordinate $q$ (resp. $\dot{q}$ and $\eta$). By following the standard procedure, the critical ``points" of $A[q]$ are obtained by imposing $D_{\eta} A[q]\!=\!0$, and once such equality should be satisfied for any admissible $\eta$, we find the well-known Euler-Lagrange (EL) equations 
\begin{align}
\frac{\partial \Lcal}{\partial q_{j}} - \frac{d}{dt}\left(\frac{\partial \Lcal}{\partial \dot{q}_{j}}\right) = 0 . \label{Ap-ELEq}
\end{align}

In classical mechanics, the above equation sets the classical trajectory followed by a system as stated by the stationary action principle, or principle of least action in Hamilton's formulation of the classical mechanics. In context of QAB, the above equation leads us to a criticality condition of the adiabaticity parameter, providing then a critical length of the adiabatic trajectory followed by the system in its {{Hilbert}} space.
The solutions of Eq.~\eqref{Ap-ELEq} allows us to predict the dynamics of the system in the phase space, but it does not mean {{that this}} trajectory is the {{optimal}} one. Such analysis is done by using the second derivative test of the functional $A[q]$ as follows
\begin{align}
D_{\eta}^2 A[q] = D_{\eta} \left( D_{\eta} A[q] \right)= \left. \frac{d \left(D_{\eta} A[q+\epsilon\eta]\right)}{d\epsilon} \right\vert_{\epsilon=0} .
\end{align}

From Eq.~\eqref{Ap-ELEq}, it is important {{to highlight}} here that the parameter $\eta$ does not {{play}} any role in the system dynamics. However, {{we will now show}} how such parameter is relevant when we compute the criticality of $A[q]$. To this {{end}}, we {{expand the above equation, which leads to}} 
\begin{align}
D_{\eta}^2 A[q] = \sum_{j=1}^{n}\int_{t_0}^{\tau} \left[ \frac{\partial^2 \Lcal}{\partial q_{j}^2} \eta^{2}_{j} + 2 \frac{\partial^2 \Lcal}{\partial q_{j}\dot{q}_{j}} \eta_{j}\dot{\eta}_{j} + \frac{\partial^2 \Lcal}{\partial \dot{q}_{j}^2} \dot{\eta}^{2}_{j} \right]dt ,
\end{align}
where we find {{a}} new condition on the parameter $\eta$ as $\dot{\eta}(t_0)\!=\!\dot{\eta}(\tau)\!=\!0$ needs to be satisfied. Now, we remark that the value of $D_{\eta}^2 A[q]$ is constrained to a suitable choice of the parameter $\eta$. Due to the evident influence of $\eta$ on the criticality of $A[q]$, we define $\Acal\!=\!\{\eta:[t_0,\tau] \rightarrow \Rmath^{n} , \eta(t_0)\!=\!\eta(\tau)\!=\!0 \}$ as the set of admissible $\eta$'s in the criticality study for $A[q]$. Therefore, given the solution $q_{\text{sol}}$ of the Eq.~\eqref{Ap-ELEq}, the Taylor's formula allows us to see that:
\begin{itemize}
	\item[1)] if $D_{\eta}^2 A[q_{\text{sol}}]>0$ for any admissible $\eta$, then the solution of $q_{\text{sol}}$ is (at least) a local minimum of $A[q_{\text{sol}}]$
	\item[2)] if $D_{\eta}^2 A[q_{\text{sol}}]<0$ for any admissible $\eta$, then the solution of $q_{\text{sol}}$ is (at least) a local maximum of $A[q_{\text{sol}}]$
	\item[3)] if there are admissible $\eta_{1}$ and $\eta_{2}$, so that $D_{\eta}^2 A[q_{\text{sol}}]<0$ for $\eta_{1}$ and $A[q_{\text{sol}}]>0$ for $\eta_{2}$, then the solution of $q_{\text{sol}}$ is a ``saddle point" of $A[q]$.
\end{itemize}

In general, the analysis on all possible values of $\eta$ is a hard task, but we can simplify the process by using its properties. For example, the boundary conditions $\eta(t_0)\!=\!\eta(\tau)\!=\!0$ implies that components of $\eta$ can be written as periodic functions with semi-period $\tau$ and, therefore, it can be described in terms of the Fourier series
\begin{align}
\eta_{n} = \sum_{n=1}^{\infty} \alpha_n \sin \left( n \pi s \right) , ~~ s = t/\tau \in [0,1] .
\end{align}

\subsection{Second Derivative test application}

Here we study the aspects of critically of the quantum adiabatic brachistochrone solution. To this end, the analysis is done by using the second derivative test of the functional $A[q]$ as follows
\begin{align}
D_{\eta}^2 A[q] = D_{\eta} \left( D_{\eta} A[q] \right)= \left. \frac{d \left(D_{\eta} A[q+\epsilon\eta]\right)}{d\epsilon} \right\vert_{\epsilon=0} .
\end{align}

It is important {{to highlight}} here that the parameter $\eta$ does not develop any role in the system dynamics. However, now we will show how such parameter is relevant when we compute the critically of $A[q]$. To this, we develop the above equation and it is possible to show that
\begin{align}
D_{\eta}^2 A[q] = \sum_{j=1}^{n}\int_{t_0}^{\tau} \left[ \frac{\partial^2 \Lcal}{\partial q_{j}^2} \eta^{2}_{j} + 2 \frac{\partial^2 \Lcal}{\partial q_{j}\dot{q}_{j}} \eta_{j}\dot{\eta}_{j} + \frac{\partial^2 \Lcal}{\partial \dot{q}_{j}^2} \dot{\eta}^{2}_{j} \right] d t ,
\end{align}
where {{we then}} find new conditions on the parameter $\eta$ as $\dot{\eta}(t_0)\!=\!\dot{\eta}(\tau)\!=\!0$, which need to be satisfied. Now, we remark that the value of $D_{\eta}^2 A[q]$ is constrained to a suitable choice of the parameter $\eta$. Due to the evident influence of $\eta$ on the critically of $A[q]$, we define 
\begin{align}
\Acal=\{\eta:[t_0,\tau] \rightarrow \Rmath^{n} , \eta(t_0)=\eta(\tau)=\dot{\eta}(t_0)=\dot{\eta}(\tau)=0 \} ,
\end{align}
as the set of admissible $\eta$'s in the criticality study for $A[q]$. Then, we have
\begin{align}
D_{\eta}^2 A[q] = \int_{0}^{\tau} \sum_{j=1}^{2}&\left[\Xi_{1}^{(j)}(t) \dot{\eta}^{2}_{j}(t) + \Xi_{2}^{(j)}(t) \dot{\eta}_{j}(t)\eta_{j}(t)\right. \nonumber\\&+ \left.\Xi_{3}^{(j)}(t) \eta^{2}_{j}(t)\right] d t ,
\end{align}
where, 
\begin{align}
\Xi_{1}^{(j)}(t) = \frac{\partial^2 \Lcal}{\partial \Omega_{j}^2} , ~~
\Xi_{2}^{(j)}(t) = 2\frac{\partial^2 \Lcal}{\partial \Omega_{j}\partial \dot{\Omega}_{j}} ,
~~ \Xi_{3}^{(j)}(t) = \frac{\partial^2 \Lcal}{\partial \dot{\Omega}^{2}_{j}} . \label{Ap-Xis}
\end{align}
Now, given the boundary conditions on the function $\eta(t)$, we can write an arbitrary function $\eta(t)$ as
\begin{align}
\eta_{j}(t) = \sum_{n=1}^{\infty} \alpha_n^{(j)} \sin \left(  \frac{n \pi t}{\tau} \right) , ~~ t \in [0,\tau] ,
\end{align}
for arbitrary real numbers $\alpha_n$, so that
\begin{align}
\dot{\eta}_{j}(t) = \sum_{n=1}^{\infty} n \pi \alpha_n^{(j)} \cos \left(  \frac{n \pi t}{\tau} \right) , ~~ t \in [0,\tau] .
\end{align}

By using these two equations, we can rewrite $D_{\eta}^2 A[q]$ as
\begin{align}
D_{\eta}^2 A[q] = \sum_{j=1}^{2} \Lambda_{j},
\end{align}
in which
\begin{align}
\Lambda_{j} &= \sum_{n,k=1}^{\infty} \alpha_n^{(j)} \alpha_k^{(j)} \int_{0}^{\tau} \Xi_{1}^{(j)}(t) \sin \left(  \frac{n \pi t}{\tau} \right)\sin \left(  \frac{k \pi t}{\tau} \right) d t \nonumber \\
&+ \sum_{n,k=1}^{\infty} n \pi \alpha_n^{(j)} \alpha_k^{(j)} \int_{0}^{\tau} \Xi_{2}^{(j)}(t)  \cos \left(  \frac{n \pi t}{\tau} \right) \sin \left(  \frac{k \pi t}{\tau} \right) d t\nonumber \\
&+ \sum_{n,k=1}^{\infty} k n \pi^2 \alpha_n^{(j)} \alpha_k^{(j)} \int_{0}^{\tau} \Xi_{3}^{(j)}(t)  \cos \left(  \frac{n \pi t}{\tau} \right) \cos \left(  \frac{k \pi t}{\tau} \right) d t .
\end{align}

Then, from above equation we can study the criticality of the brachistochrone solutions {considered here}.

\subsection{Criticality of the braquistochrone in Eq.~(\ref{Ap-eq:quadratic_sol})}

By using the Eq.~(\ref{Ap-eq:quadratic_sol}) in Eqs.~\eqref{Ap-Xis}, we find that
\begin{align}
\Xi_{1}^{(j)}(t) &= \pi^2 \left[2 + 3 (-1)^j \cos\left(  \frac{\pi t}{\tau} \right)\right] , ~
\Xi_{2}^{(j)}(t) = 4\pi (-1)^j \sin\left(  \frac{\pi t}{\tau} \right) , \nonumber
\end{align}
and $\Xi_{3}^{(j)}(t)\!=\!2$. Then, we have
\begin{align}
\Lambda_{j} &= \sum_{n,k=1}^{\infty} \alpha_n^{(j)} \alpha_k^{(j)} \pi^2 \delta_{nk} + \sum_{n,k = 1}^{\infty} k n \pi^2 \alpha_n^{(j)} \alpha_k^{(j)} \delta_{nk} \nonumber \\
&= \sum_{n}^{\infty} \left(\alpha_n^{(j)}\pi\right)^2 (1+n^2) .
\end{align}

Therefore, we conclude that 
\begin{align}
D_{\eta}^2 A[q] &= \sum_{n}^{\infty} \left(\alpha_n^{(j)}\pi\right)^2 (1+n^2) \geq 0 , ~~ \forall\; \alpha_n \in \Rmath .
\end{align}
{{This proves}} that the Brachistochrone given in Eq.~(\ref{Ap-eq:quadratic_sol}) is a minimum of the functional adiabatic time.

\section{The Quantum Battery Charging process} 

In our work, the quantum battery is represented by the {{three-level}} Hamiltonian $H_0 = \sum_{n=1}^{2}\hbar\omega_{(n-1)n}\ketbra{\epsilon_i}{\epsilon_i}$, with $\omega_{01}<\omega_{12}<\omega_{23}$~\cite{Santos:19-a}. 
The charging process is expressed as a time-dependent Hamiltonian that will drive $H_0$ from the ground state, $\ket{\epsilon_0}$, to the second excited one, $\ket{\epsilon_2}$, 
\begin{equation}
H(t) = \hbar\Omega_{1}(t)\ketbra{\epsilon_0}{\epsilon_1} + \hbar\Omega_{2}(t)\ketbra{\epsilon_1}{\epsilon_2} + h.c.  ,
\end{equation}
where $\Omega_{1}$ and $\Omega_{2}$ are identified as the generalized coordinates.
{{Since they}} represent the physical Rabi frequencies, {{they}} will be considered real functions. 
Note that we are considering resonance between $\Omega_i$ and the energy gaps of the system. 

The eigenvectors of $H(t)$ are 
\begin{align}
\ket{E_\pm(t)} &= \frac{1}{\sqrt{2}}\left[\frac{\Omega_{1}(t)}{\Delta(t)}\ket{\epsilon_0}  \pm \ket{\epsilon_1} + \frac{\Omega_{2}(t)}{\Delta(t)}\ket{\epsilon_2}\right] ,\\
\ket{E_0(t)} &= \frac{\Omega_{2}(t)}{\Delta(t)}\ket{\epsilon_0} - \frac{\Omega_{2}(t)}{\Delta(t)}\ket{\epsilon_2} , 
\end{align}
with the energies $E_{\pm}(t)=\pm\Delta(t)$ and $E_0(t)=0$, with $\Delta^2(t)=\Omega_{1}^2(t) + \Omega_{2}^2(t)$ also being the minimum energy gap between subsequent eigenvectors.
We call $\ket{E_\pm}$ \textit{bright} states and $\ket{E_0}$ \textit{dark} state. 

Moreover, one can perform the adiabatic passage from the $\ket{\epsilon_0}$ to $\ket{\epsilon_2}$ in {{two}} distinct ways. 
Firstly, by setting the boundary conditions, $\left\{\Omega_{1}(0), \Omega_2(\tau) \right\}\neq 0$ and $\left\{\Omega_{1}(\tau),\Omega_{2}(0)\right\} = 0$, the initial state $\ket{\epsilon_0}$ is written as a symmetric superposition of the bright states,
\begin{equation}\label{Ap-eq:bright_passage}
\ket{\Psi(0)} = \ket{\epsilon_0} =\frac{\ket{E_+(0)} + \ket{E_-(0)}}{\sqrt{2}}   .
\end{equation}
Secondly, we can the set  $\left\{\Omega_{1}(\tau), \Omega_2(0) \right\}\neq 0$ and $\left\{\Omega_{1}(0),\Omega_{2}(\tau)\right\} = 0$, thereby having the ground state expressed as the \textit{dark} state, 
\begin{equation}\label{Ap-eq:dark_passage}
\ket{\Psi(0)} = \ket{\epsilon_0} = \ket{E_0(0)}   .
\end{equation}
{In the main text,} the population inversion {{process}} through {above equation is refereed as} stable charging process, and through Eq.~(\ref{Ap-eq:bright_passage}) {refers to an} unstable charging process. {{Let us examine}} by looking at the system ergotropy.

{{First, we}} calculate the ergotropy {{when performing}} an adiabatic evolution using the Eq.~(\ref{Ap-eq:bright_passage}).
By applying the Schr\"{o}dinger evolution {{we find}} that the evolved state is 
\begin{equation}
\begin{split}
\ket{\psi^{\textnormal{ad}}_{\textnormal{bright}}(t)} &= \frac{1}{\sqrt{2}}\left\{e^{-i\int E_+(t')d t'}\ket{E_+(t)} + e^{-i\int E_-(t')d t'}\ket{E_-(t)} \right\} \\
&=  \frac{e^{-i\int E_+(t')d t'}}{\sqrt{2}}\left[ \frac{\Omega_{1}(t)}{\Delta(t)}\ket{\epsilon_0} + \ket{\epsilon_1} +  \frac{\Omega_{2}(t)}{\Delta(t)}\ket{\epsilon_2}\right] \\
&+ \frac{e^{-i\int E_-(t')d t'}}{\sqrt{2}} \left[ \frac{\Omega_{1}(t)}{\Delta(t)}\ket{\epsilon_0} - \ket{\epsilon_1} +  \frac{\Omega_{2}(t)}{\Delta(t)}\ket{\epsilon_2}\right]  ,
\end{split}
\end{equation}
and, regrouping the terms, 
\begin{equation}
\begin{split}
\ket{\psi^{\textnormal{ad}}_{\textnormal{bright}}(t)} &= \frac{\cos\left(\tilde{\theta}(t)\right)}{\Delta(t)} \left[\Omega_{1}(t)\ket{\epsilon_0} +  \Omega_{2}(t)\ket{\epsilon_2}\right] \\
&- i\sin\left(\tilde{\theta}(t) \right)\ket{\epsilon_1} ,
\end{split}
\end{equation}
where $\tilde{\theta}(t)=\int_0^t \Delta(t') dt'$, and we used that our system satisfies the parallel transport condition, $\interpro{E_n(t)}{\dot{E}_n(t)}=0$. 
Further, the system ergotropy when performing the \textit{bright} passage is
\begin{equation}\label{Ap-eq:ergotropy_unstable}
\begin{split}
C(t) =  &\mel{\psi^\textnormal{ad}(t)}{H_0}{\psi^\textnormal{ad}(t)} - \mel{\epsilon_0}{H_0}{\epsilon_0} \\
&=\frac{\cos^2\left(\tilde{\theta}(t)\right)}{\Delta^2(t)} \left[\omega_{01}\Omega_{1}^2(t) + \omega_{12}\Omega_{2}^2(t)\right] + \sin^2\left(\tilde{\theta}(t) \right) - \omega_{01}  .
\end{split}
\end{equation}
From Eq.~(\ref{Ap-eq:ergotropy_unstable}) we can see that the stored energy has an oscillating term, which means that the battery will not {{remain}} fully charged after the times $t > \tau$.

On the other hand, we can use the stimulated Raman adiabatic passage protocol in order to perform the population transfer. 
Considering an adiabatic evolution stating with the dark state, Eq.~(\ref{Ap-eq:dark_passage}), we have the evolved state \begin{equation}
\ket{\Psi^{\textnormal{ad}}(t)} = \ket{E_0(t)} =  \frac{\Omega_2(t)}{\Delta(t)}\ket{\epsilon_0} - \frac{\Omega_1(t)}{\Delta(t)}\ket{\epsilon_2}  ,
\end{equation}
where $\interpro{E_0(t)}{\dot{E}_0(t)}=0$.
Hence, the system ergotropy for this scenario reads 
\begin{equation}
\begin{split}
C(t) &= \bra{\psi^\textnormal{ad}_{\textnormal{dark}}(t)}{H_0}\ket{\psi^\textnormal{ad}_{\textnormal{dark}}(t)} - \mel{\epsilon_i}{H_0}{\epsilon_1} \\
&= \frac{\omega_{12}\Omega_{1}^2(t) + \omega_{01}\Omega_{2}^2(t)}{{\Delta^2(t)}}  - \omega_{01}  .
\end{split}
\end{equation}
Therefore, when the system attains its maximum charge it will keep its charge.

We interpret the unstable behavior of the \textit{bright} passage as relative to the destructive and constructive superpositions introduced by the complex phases $e^{-i\int_0^t E\pm(t')dt'}$.

\section{Self-discharge process}

{{The decay}} dynamics of the system is given by
\begin{equation}\label{Ap-LindEq}
\dot{\rho}(t) = \frac{1}{i\hbar} [H(t),\rho(t)] + \Lcal_{\text{rel}}[\rho(t)]+ \Lcal_{\text{dep}}[\rho(t)]  , 
\end{equation}
where the superoperators $\Lcal_{\text{rel}}[\bullet]$ and $\Lcal_{\text{dep}}[\bullet]$ describe the relaxation and dephasing phenomena, respectively, {{which}} can be written as
\begin{subequations}\label{Ap-eq:RelTerm}
	\begin{align}
	\Lcal_{\text{rel}}[\bullet] &= \sum_{k\neq j}\Gamma_{kj} \left[\sigma_{kj}\bullet\sigma_{jk} - \frac{1}{2}\{\sigma_{kk},\bullet\} \right]  , \\
	\Lcal_{\text{dep}}[\bullet] &= \sum_{j=2,3}\gamma_{j} \left[\sigma_{jj}\bullet\sigma_{jj} - \frac{1}{2}\{\sigma_{jj},\bullet\} \right]  ,
	\end{align}
\end{subequations}
where $\sigma_{kj}\!=\!|\varepsilon_{k}\rangle\langle \varepsilon_{j}|$ and $\Gamma_{kj}\!=\!\Gamma_{jk}$. Building on the general definitions we have introduced in Eqs.~(\ref{Ap-eq:RelTerm}), we would like to clarify two important points on the characteristics of noise we consider in the rest of this work. First, the relaxation processes we consider are only the sequential decays, meaning, $\ket{\varepsilon_{3}}\! \rightarrow \!\ket{\varepsilon_{2}}$ and $\ket{\varepsilon_{2}} \!\rightarrow\! \ket{\varepsilon_{1}}$ characterized by the rates $\Gamma_{32}$ and $\Gamma_{21}$, respectively. Then, by writing $\varrho_{nm}(t)$ as the instantaneous matrix elements of $\rho(t)$, we obtain the set of differential equations for the diagonal elements as
\begin{equation}
\dot{\varrho}_{11}(t) = -\Gamma_{10}\varrho_{11}(t) + \Gamma_{21}\varrho_{22}(t),~~
\dot{\varrho}_{22}(t) = - \Gamma_{21}\varrho_{22}(t)  . \label{Ap-Dif}
\end{equation}
and for the off-diagonal elements as
\begin{align}
\dot{\varrho}_{nm,n\neq m}(t) = z_{nm} \varrho_{nm,n\neq m}(t) ,
\end{align}
where $z_{nm}$ is a complex number. As a first remark, we notice that for initial states in which $\dot{\varrho}_{nm,n\neq m}(0)\!=\!0$, then we have the solution $\dot{\varrho}_{nm,n\neq m}(t)\!=\!0$. so that we just need to solve the set of equations given in Eq.~\eqref{Ap-Dif}.
We can use the Laplace transform to solve the above equations. By denoting $\chi_{nn}(s)$ as the Laplace transformation of $\varrho_{nn}(t)$, we find the system of linear equations given by
\begin{align}
s\chi_{11}(s) &= -\Gamma_{10}\chi_{11}(s) + \Gamma_{21}\chi_{22}(s) + \varrho_{11}(0) , \nonumber \\
s\chi_{22}(s) &= - \Gamma_{21}\chi_{22}(s) + \varrho_{22}(0)  .
\end{align}
Now, we can isolate the Laplace variables $\chi_{nn}$ and find the equations,
\begin{equation}
\begin{split}
\chi_{11}(s) &= \frac{(s+\Gamma_{21})\varrho_{11}(0)+\Gamma_{21}\varrho_{22}(0)}{(s+\Gamma_{21})(s+\Gamma_{10})} , \\
\chi_{22}(s) &= \frac{\varrho_{22}(0)}{s + \Gamma_{21}}  .
\end{split}
\end{equation}
Finally, we use the inverse transformation and get
\begin{equation}
\begin{split}
\varrho_{11}(t) &= \frac{1}{\Gamma_{10} - \Gamma_{21}}\left\{e^{-t\Gamma_{21}}\Gamma_{21}\varrho_{22}(0) \right.\\
&\left. + e^{-t\Gamma_{10}}\left[ \Gamma_{10}\varrho_{11}(0) - \Gamma_{21} \left(\varrho_{11}(0)+\varrho_{22}(0)\right)\right] \right\}, \\
\varrho_{22}(t) &= e^{-t\Gamma_{21}}\varrho_{22}(0)  .
\end{split}
\end{equation}
Therefore, by using the case in which the battery is fully charged initially, where the initial conditions are $\varrho_{00}(0)=\varrho_{11}(0)=0$ and $\varrho_{22}(0)=1$, one gets
\begin{subequations}\label{Ap-eq:selfDischarge_solutions}
	\begin{align}
	\varrho_{00}(t) &= 1 - \frac{\Gamma_{10}e^{-t\Gamma_{21}} - \Gamma_{21}e^{-t\Gamma_{10}}}{\Gamma_{10}-\Gamma_{21}} , \\
	\varrho_{11}(t) &= \frac{\Gamma_{21}\left(e^{-t\Gamma_{21}} - e^{-t\Gamma_{10}}\right)}{\Gamma_{10}-\Gamma_{21}} , \\
	\varrho_{22}(t) &= e^{-t\Gamma_{21}}  .
	\end{align}
\end{subequations}

The amount of charge stored in the battery is measured by the ergotropy, which is defined as
\begin{align}
\Ecal(t) = \sum\nolimits_{i,n}^{N,N} \varrho_{n}(t) \epsilon_{i} \left( |\interpro{\varrho_{n}(t)}{i}|^2 - \delta_{ni} \right) , \label{Ap-ErgotropyXstates}
\end{align}	
with the ordered set of eigenenergies $\epsilon_{1}\!\leq\!\epsilon_{2}\!\leq\!\cdots\!\leq\!\epsilon_{N}$ with eigenstates $\ket{n}$, of the internal battery Hamiltonian $H_0$ and the instantaneous spectral decomposition $\varrho_{1}(t)\!\geq\!\varrho_{2}(t)\!\geq\!\cdots\!\geq\!\varrho_{N}(t)$ of the instantaneous battery state $\rho(t)$, associated to eigenvectors $\ket{\varrho_{n}(t)}$. Therefore, by using that $\varrho_0 < \varrho_1 < \varrho_2$, we can see that
\begin{align}
\Ecal(t) &= \left\{ \begin{matrix}
\varepsilon_1 ( \varrho_1 - \varrho_0 ) + \varepsilon_2 ( \varrho_2 - \varrho_1 ) & \text{if}~\varrho_2\!>\!\varrho_1\!>\!\varrho_0 \\
( \varrho_2 - \varrho_1 )(\varepsilon_2 - \varepsilon_1 ) & \text{if}~ \varrho_2\!>\!\varrho_0\!\geq\!\varrho_1 \\
\varepsilon_2 ( \varrho_2 - \varrho_0 )& \text{if}~ \varrho_0\!>\!\varrho_2\!>\!\varrho_1 \\
0 & \text{if}~ \varrho_0\!>\!\varrho_1\!>\!\varrho_2
\end{matrix} \right. ,
\end{align}
where we simplify the notation $\varrho_n\!=\!\varrho_{nn}(t)$.


\end{document}